\documentclass[conference]{IEEEtran}
\IEEEoverridecommandlockouts %

\usepackage{cite} %
\usepackage{graphicx} %
\usepackage{amsmath,amsfonts,amsbsy,amssymb,amsthm}
\usepackage[caption=false,font=scriptsize]{subfig} %

\def\headertext{~\hfill{\scriptsize\thepage}}
\def\footertext{~\hfill~}

\makeatletter
\def\ps@headings{%
\def\@oddhead{\headertext}%
\def\@evenhead{\headertext}%
\def\@oddfoot{\footertext}%
\def\@evenfoot{\footertext}}
\usepackage[scaled=0.78]{beramono}

\DeclareSymbolFont{bbold}{U}{bbold}{m}{n}
\DeclareSymbolFontAlphabet{\mathbbold}{bbold}

\newcommand{\ceil}[1]{{\left\lceil{#1}\right\rceil}}

\newcommand{\EEE}{{\mathcal{E}}}
\newcommand{\floor}[1]{{\left\lfloor{#1}\right\rfloor}}
\newcommand{\m}[1]{{\mbox{#1}}}
\newcommand{\vv}{{\mathbf{v}}}
\newcommand{\yy}{{\mathbf{y}}}
\newcommand{\ZZ}{{\mathbb{Z}}}

\newcommand{\mCW}{{\textup{\textsf{\tiny{CW}}}}}
\newcommand{\mSW}{{\textup{\textsf{\tiny{SW}}}}}
\newcommand{\mB}{{\textup{\textsf{\tiny{B}}}}}

\allowdisplaybreaks

\theoremstyle{definition}
\newtheorem*{thm:example}{Example}
\newtheorem*{thm:problem}{Problem}
\newtheorem*{thm:definition}{Definition}
\newtheorem*{thm:conjecture}{Conjecture}

\theoremstyle{plain}
\newtheorem{thm:claim}{Claim}
\newtheorem{thm:proposition}{Proposition}
\newtheorem{thm:lemma}{Lemma}
\newtheorem{thm:corollary}{Corollary}
\newtheorem{thm:theorem}{Theorem}

\begin{document}

\title{Erasure Coding for Real-Time Streaming}

\author{%
\IEEEauthorblockN{Derek Leong}
\IEEEauthorblockA{Department of Electrical Engineering\\
California Institute of Technology\\
Pasadena, California 91125, USA\\
\texttt{derekleong@caltech.edu}}
\and
\IEEEauthorblockN{Tracey Ho}
\IEEEauthorblockA{Department of Electrical Engineering\\
California Institute of Technology\\
Pasadena, California 91125, USA\\
\texttt{tho@caltech.edu}}%
\thanks{%
\hrule width 0.33\columnwidth \vskip5pt
This paper is an extended version of \cite{dl:leong12streaming}, which was presented at the ISIT~2012 conference.}
\thanks{%
This work was supported in part
by the Air Force Office of Scientific Research under Grant FA9550-10-1-0166.}
} %

\maketitle

\begin{abstract}
We consider a real-time streaming system where messages are created sequentially at the source, and are encoded for transmission to the receiver over a packet erasure link.
Each message must subsequently be decoded at the receiver within a given delay from its creation time.
The goal is to construct an erasure correction code that achieves the maximum message size when all messages must be decoded by their respective deadlines under a specified set of erasure patterns (erasure model).
We present an explicit intrasession code construction that is asymptotically optimal under erasure models containing a limited number of erasures per coding window, per sliding window, and containing erasure bursts of a limited length.
\end{abstract}

\section{Introduction}

We consider a real-time streaming system where messages are created sequentially at the source, and are encoded for transmission to the receiver over a packet erasure link.
Each message must subsequently be decoded at the receiver within a given delay from its creation time.
We seek to construct an erasure correction code that withstands a specified set of erasure patterns (erasure model), allowing all messages to be decoded by their respective deadlines.

In particular, we consider three erasure models:
the first model limits the number of erasures in each coding window,
the second model limits the number of erasures in each sliding window,
while the third model limits the length of erasure bursts.
For each erasure model, the objective is to find an optimal code that achieves the maximum message size, among all codes that allow all messages to be decoded by their respective deadlines under all admissible erasure patterns.

We present an explicit intrasession code construction which specifies an allocation of link bandwidth or data packet space among the different messages;
coding occurs within each message but not across messages.
Intrasession coding is attractive due to its relative simplicity, but it is not known in general when intrasession coding is sufficient or when intersession coding is necessary.
We show that for an asymptotic number of messages, our code construction achieves the optimal message size among all codes (intrasession or intersession) for the first and second models with any given maximum number of erasures per window, and for the third model when the given maximum erasure burst length is sufficiently short or long.

In related work, Martinian \textit{et al.}~\cite{dl:martinian02low, dl:martinian07delay} provided constructions of streaming codes that minimize the delay required to correct erasure bursts of a given length.
Streaming codes in which the decoding error probability decays exponentially with delay, called tree codes or anytime codes, are considered in~\cite{dl:schulman96coding, dl:sahai01anytime, dl:sukhavasi12distributed}.
Tekin \textit{et al.}~\cite{dl:tekin11erasure} considered erasure correction coding for a non-real-time streaming system where all messages are initially present at the encoder.

We begin with a formal definition of the problem in Section~\ref{sec:ProblemDefinition}, and proceed to describe the construction of our intrasession code in Section~\ref{sec:CodeConstruction}.
We then demonstrate the optimality of this code under each erasure model in the subsequent sections.
Proofs of theorems are deferred to Appendix~\ref{sec:ProofsOfTheorems}.

\section{Problem Definition}
\label{sec:ProblemDefinition}

\begin{figure}
\centering
\includegraphics[width=250pt]{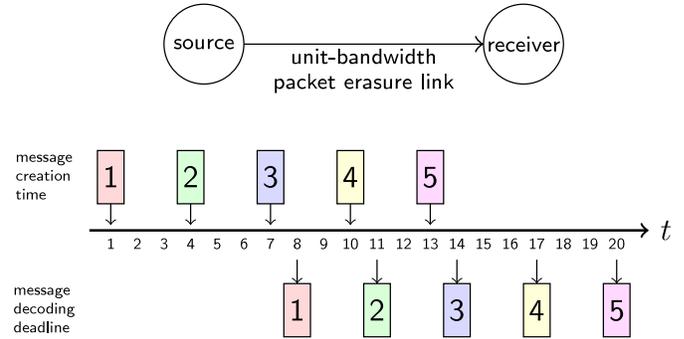}
\caption{Real-time streaming system for \m{$(c,d)=(3,8)$}.
Each of the messages $\{1,\ldots,5\}$ is assigned a unique color.
Messages are created at the source at regular intervals of $c$ time steps, and must be decoded at the receiver within $d$ time steps from their respective creation times.
At each time step~$t$, the source is allowed to transmit a single data packet of normalized unit size over the link.}
\label{fig:SystemDiagram}
\end{figure}

Consider a discrete-time data streaming system comprising a source and a receiver, with a directed unit-bandwidth packet erasure link from the source to the receiver.
Independent messages of uniform size \m{$s>0$} are created at the source at regular intervals of \m{$c\in\ZZ^+$} time steps, and must be decoded at the receiver within \m{$d\in\ZZ^+$} time steps from their respective creation times.
At each time step \m{$t\in\ZZ^+$}, the source is allowed to transmit a single data packet of normalized unit size over the link.
Fig.~\ref{fig:SystemDiagram} depicts this real-time streaming system for an instance of \m{$(c,d)$}.

More precisely, each message~\m{$k\in\ZZ^+$} is created at time step \m{$(k-1)c+1$}, and is to be decoded by time step \m{$(k-1)c+d$}.
The coded data transmitted at each time step~\m{$t\in\ZZ^+$} must be a function of messages created at time step~$t$ or earlier.
Let coding window $W_k$ be the interval of $d$ time steps between the creation time and the decoding deadline of message~$k$, i.e.,
\[
W_k
\triangleq\{(k-1)c+1,\ldots,(k-1)c+d\}.
\]
We shall assume that \m{$d>c$} so as to avoid the degenerate case of nonoverlapping coding windows for which it is sufficient to code individual messages separately.

Consider the first $n$ messages \m{$\{1,\ldots,n\}$}, and the union of their (overlapping) coding windows $T_n$ given by
\[
T_n
\triangleq W_1\cup\cdots\cup W_n
= \{1,\ldots,(n-1)c+d\}.
\]
An erasure pattern \m{$E\subseteq T_n$} specifies a set of erased data transmissions over the link.
More precisely, if \m{$t\in E$}, then none of the data transmitted at time step~$t$ is received by the receiver (i.e., the entire data packet is erased);
if \m{$t\in T_n\backslash E$}, then all of the data transmitted at time step~$t$ is received by the receiver at time step~$t$ (i.e., the entire data packet is received without delay).
An erasure model specifies a set of erasure patterns that an erasure correction code should withstand.

For a given pair of positive integers $a$ and $b$, we define the offset quotient $q_{a,b}$ and remainder $r_{a,b}$ to be the unique integers satisfying the following three conditions:
\[
a=q_{a,b}\,b+r_{a,b},\qquad
q_{a,b}\in\ZZ^+_0,\qquad
r_{a,b}\in\{1,\ldots,b\},
\]
where $\ZZ^+_0$ denotes the set of nonnegative integers, i.e., \m{$\ZZ^+\cup\{0\}$}.
Note that our definition departs from the usual definition of quotient and remainder in that $r_{a,b}$ can be equal to $b$ but not $0$.

\section{Code Construction}
\label{sec:CodeConstruction}

We present an intrasession code which codes only within each message and not across different messages.
We begin by specifying the amount of link bandwidth or data packet space allocated for the encoding of each message at each time step.
An appropriate code (e.g., random linear coding, MDS code) is then applied to the allocation so that each message can be decoded whenever the total amount of received data that encodes that message is at least the message size~$s$.

\begin{figure}
\centering
\subfloat[$(c,d)=(3,9)$]
{\includegraphics[width=250pt]{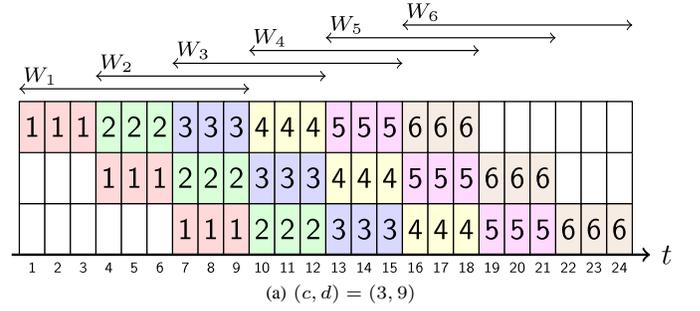}}
\\
\subfloat[$(c,d)=(3,8)$]
{\includegraphics[width=250pt]{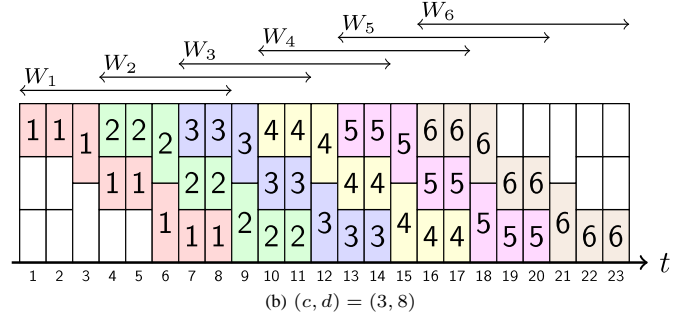}}
\caption{Allocation of link bandwidth at each time step~$t$, in the encoding of messages \m{$\{1,\ldots,6\}$}, for (a)~\m{$(c,d)=(3,9)$} and (b)~\m{$(c,d)=(3,8)$}.
Each message is assigned a unique color.
In (a), because $d$ is a multiple of $c$, we have \m{$q_{d,c}+1=3$} active messages at each time step.
In (b), because $d$ is not a multiple of $c$, we have either \m{$q_{d,c}=2$} or \m{$q_{d,c}+1=3$} active messages at each time step.}
\label{fig:CodeConstruction}
\end{figure}

The allocation of link bandwidth follows a simple rule:
the link bandwidth at each time step is divided evenly among all \emph{active} messages.
We say that message~$k$ is active at time step~$t$ if and only if $t$ falls within its coding window, i.e., \m{$t\in W_k$}.
Fig.~\ref{fig:CodeConstruction} shows how much link bandwidth at each time step is allocated to each message, for two instances of \m{$(c,d)$}.

For a given choice of \m{$(c,d)$}, the messages that are encoded at a given time step~\m{$t\in\ZZ^+$} can be stated explicitly as follows:
First, we define $A_t$ to be the set of active messages at time step~$t$, i.e.,
\begin{align*}
A_t
&\triangleq \{k\in\ZZ^+:t\in W_k\}
\\
&= \{k\in\ZZ^+:(k-1)c+1\leq t\leq(k-1)c+d\}
\\
&= \left\{k\in\ZZ^+:\frac{t-d}{c}+1\leq k\leq\frac{t-1}{c}+1\right\}.
\end{align*}
Treating nonpositive messages $0,-1,-2,\ldots$ as dummy messages, we can write
\[
A_t
= \left\{\ceil{\frac{t-d}{c}+1},\ldots,\floor{\frac{t-1}{c}+1}\right\}.
\]
Expressing this in terms of $q_{d,c}$, $r_{d,c}$, $q_{t,c}$, $r_{t,c}$ yields
\[
A_t
= \left\{q_{t,c}+1-q_{d,c}+\ceil{\frac{r_{t,c}-r_{d,c}}{c}},\ldots,q_{t,c}+1\right\}.
\]
It follows that the number of active messages $|A_t|$ varies over time depending on the value of $r_{t,c}$;
specifically, two cases are possible:

\m{\textit{Case 1}:}
If \m{$r_{t,c}\leq r_{d,c}$}, then
\[
-1<\frac{1-c}{c}\leq\frac{r_{t,c}-r_{d,c}}{c}\leq 0,
\]
which implies that \m{$\ceil{\frac{r_{t,c}-r_{d,c}}{c}}=0$}, and
\[
A_t
= \left\{q_{t,c}+1-q_{d,c},\ldots,q_{t,c}+1\right\}.
\]
The \m{$q_{d,c}+1$} messages of $A_t$ are therefore encoded at time step~$t$, with each message allocated $\frac{1}{q_{d,c}+1}$ amount of link bandwidth.

\m{\textit{Case 2}:}
If \m{$r_{t,c}>r_{d,c}$}, then
\[
0<\frac{r_{t,c}-r_{d,c}}{c}\leq\frac{c-1}{c}<1,
\]
which implies that \m{$\ceil{\frac{r_{t,c}-r_{d,c}}{c}}=1$}, and
\[
A_t
= \left\{q_{t,c}+1-(q_{d,c}-1),\ldots,q_{t,c}+1\right\}.
\]
The \m{$q_{d,c}$} messages of $A_t$ are therefore encoded at time step~$t$, with each message allocated $\frac{1}{q_{d,c}}$ amount of link bandwidth.

Note that when $d$ is a multiple of $c$, we have \m{$r_{t,c}\leq r_{d,c}=c$} for any $t$, which implies that \m{$q_{d,c}+1$} messages are encoded at every time step.

In our subsequent performance analysis of this code, we make repeated use of two key code properties;
these are presented as technical lemmas in Appendix~\ref{sec:CodeProperties}.

\section{Performance under $z$ Erasures\\
per Coding Window}
\label{sec:CodingWindow}

For the first erasure model, we look at erasure patterns that have a limited number of erasures per coding window.
Consider the first $n$ messages \m{$\{1,\ldots,n\}$}, and the union of their (overlapping) coding windows $T_n$.
Let $\EEE_n^\mCW$ be the set of erasure patterns that have $z$ or fewer erasures in each coding window $W_k$, i.e.,
\[
\EEE_n^\mCW
\triangleq \big\{
E\subseteq T_n:
|E\cap W_k|\leq z \;\forall\; k\in\{1,\ldots,n\}
\big\}.
\]
The objective is to construct a code that allows all $n$ messages \m{$\{1,\ldots,n\}$} to be decoded by their respective deadlines under any erasure pattern \m{$E\in\EEE_n^\mCW$}.
Let $s_n^\mCW$ be the maximum message size that can be achieved by such a code, for a given choice of \m{$(n,c,d,z)$}.

We observe that over a finite time horizon (i.e., when the number of messages $n$ is finite), intrasession coding can be suboptimal.
The following example shows that an intersession code can achieve a message size that is strictly larger than the message size achieved by an optimal intrasession code:

\begin{thm:example}[Finite time horizon]
Suppose that \m{$(n,c,d,z)$} $=$ \m{$(3,1,3,1)$}.
The maximum message size that can be achieved by an intrasession code is $\frac{6}{7}$;
one such optimal intrasession code, which can be found by solving a linear program, is as follows (the amount of link bandwidth allocated to each message is indicated in parentheses):

\begin{center}
\includegraphics[width=110pt]{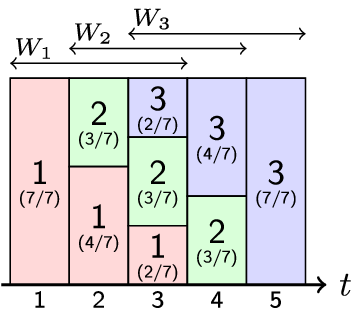}
\end{center}

\noindent
The following intersession code achieves a strictly larger message size of $1$ ($M_k$ denotes message~$k$):

\begin{center}
\includegraphics[width=110pt]{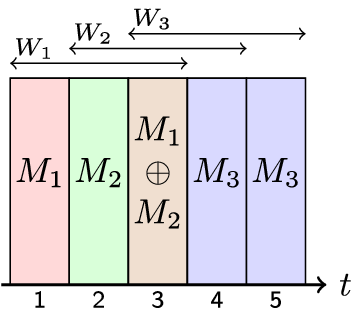}
\end{center}

\noindent
Using a simple cut-set bound argument, we can show that this is also the maximum achievable message size, i.e., $s_n^\mCW=1$.
\end{thm:example}

However, it turns out that the intrasession code constructed in Section~\ref{sec:CodeConstruction} is \emph{asymptotically} optimal;
the gap between the maximum achievable message size $s_n^\mCW$ and the message size achieved by the code vanishes as the number of messages $n$ goes to infinity:

\begin{thm:theorem}
\label{thm:theorem:CodingWindowOptimalCode}
The code constructed in Section~\ref{sec:CodeConstruction} is asymptotically optimal in the following sense:
the code achieves a message size of
\[
\sum_{j=1}^{d-z} y_j,
\]
which is equal to the asymptotic maximum achievable message size
$\lim_{n\rightarrow\infty}s_n^\mCW$,
where \m{$\yy=(y_1,\ldots,y_d)$} is defined as
\[
\yy
\triangleq \bigg(\overbrace{
\underbrace{\frac{1}{q_{d,c}+1},\ldots,\frac{1}{q_{d,c}+1}}_{(q_{d,c}+1)r_{d,c}\text{ entries}},
\underbrace{\frac{1}{q_{d,c}},\ldots,\frac{1}{q_{d,c}}}_{q_{d,c}(c-r_{d,c})\text{ entries}}
}^{d\text{ entries}}\bigg).
\]
\end{thm:theorem}

The achievability claim of this theorem is a consequence of Lemma~\ref{thm:lemma:Achievability};
to prove the converse claim, we consider a cut-set bound corresponding to a specific \emph{worst-case} erasure pattern in which exactly $z$ erasures occur in every coding window.
This erasure pattern is chosen with the help of Lemma~\ref{thm:lemma:PartitionCodingWindows};
specifically, the erased time steps are chosen to coincide with the larger blocks allocated to each message in the constructed code.

\section{Performance under $z$ Erasures\\
per Sliding Window of $d$ Time Steps}
\label{sec:SlidingWindow}

For the second erasure model, we look at erasure patterns that have a limited number of erasures per sliding window of $d$ time steps.
Consider the first $n$ messages \m{$\{1,\ldots,n\}$}, and the union of their (overlapping) coding windows $T_n$.
Let sliding window $L_t$ denote the interval of $d$ time steps beginning at time step~$t$, i.e.,
\[
L_t
\triangleq \{t,\ldots,t+d-1\}.
\]
Let $\EEE_n^\mSW$ be the set of erasure patterns that have $z$ or fewer erasures in each sliding window $L_t$, i.e.,
\[
\EEE_n^\mSW
\triangleq \big\{
E\subseteq T_n:
|E\cap L_t|\leq z \;\forall\; t\in\{1,\ldots,(n-1)c+1\}
\big\}.
\]
The objective is to construct a code that allows all $n$ messages \m{$\{1,\ldots,n\}$} to be decoded by their respective deadlines under any erasure pattern \m{$E\in\EEE_n^\mSW$}.
Let $s_n^\mSW$ be the maximum message size that can be achieved by such a code, for a given choice of \m{$(n,c,d,z)$}.

We note that since \m{$\EEE_n^\mSW\subseteq\EEE_n^\mCW$}, we therefore have \m{$s_n^\mSW\geq s_n^\mCW$}.
For the special case of \m{$c=1$}, each sliding window is also a coding window, and so this sliding window erasure model reduces to the coding window erasure model of Section~\ref{sec:CodingWindow}, i.e., \m{$\EEE_n^\mSW=\EEE_n^\mCW$}.
Over a finite time horizon, intrasession coding can also be suboptimal for this erasure model;
the illustrating example from Section~\ref{sec:CodingWindow} applies here as well.

Surprisingly, the constructed intrasession code also turns out to be asymptotically optimal over all codes;
the omission of erasure patterns in $\EEE_n^\mSW$ compared to $\EEE_n^\mCW$ has not led to an increase in the maximum achievable message size (cf.~Theorem~\ref{thm:theorem:CodingWindowOptimalCode}):

\begin{thm:theorem}
\label{thm:theorem:SlidingWindowOptimalCode}
The code constructed in Section~\ref{sec:CodeConstruction} is asymptotically optimal in the following sense:
the code achieves a message size of
\[
\sum_{j=1}^{d-z} y_j,
\]
which is equal to the asymptotic maximum achievable message size
$\lim_{n\rightarrow\infty} s_n^\mSW$.
\end{thm:theorem}

Proving the converse claim of this theorem requires a different approach from that of Theorem~\ref{thm:theorem:CodingWindowOptimalCode}.
When $d$ is a multiple of $c$, we need only consider a cut-set bound corresponding to an obvious \emph{worst-case} erasure pattern in which exactly $z$ erasures occur in every sliding window, specifically, a periodic erasure pattern with alternating intervals of $z$ erased time steps and \m{$d-z$} unerased time steps.
When $d$ is not a multiple of $c$, no single admissible erasure pattern provides a cut-set bound that matches the constructed code;
instead, we need to combine different erasure patterns for different messages.
To pick these erasure patterns, we first choose a specific \emph{base} erasure pattern~$E'$ (which may not be admissible in general) with the help of Lemma~\ref{thm:lemma:PartitionCodingWindows}.
We then derive admissible erasure patterns from $E'$ by taking its intersection with each coding window, i.e., \m{$(E'\cap W_k)\in\EEE_n^\mSW$}.
These derived erasure patterns are used in the inductive computation of an upper bound for the conditional entropy
\[
H\Big(
X[W_n\backslash E']
\,\Big|\,
M_1^n,
X_1^{(n-1)c}
\Big),
\]
where $X_t$ is a random variable representing the coded data transmitted at time step~$t$,
$M_k$ is a random variable representing message~$k$, and
$X[A]\triangleq(X_t)_{t\in A}$.
Intuitively, this conditional entropy term expresses how much space is left in the unerased data packets of the coding window for message~$n$, after encoding the first $n$ messages, and conditioned on the previous time steps.
The nonnegativity of the conditional entropy leads us to a bound for $s_n^\mSW$ that matches the message size achieved by the constructed code in the limit \m{$n\rightarrow\infty$}.

\section{Performance under Erasure Bursts\\
of $z$ Time Steps}

For the third erasure model, we look at erasure patterns that contain erasure bursts of a limited number of time steps.
Consider the first $n$ messages \m{$\{1,\ldots,n\}$}, and the union of their (overlapping) coding windows $T_n$.
Let $\EEE_n^\mB$ be the set of erasure patterns in which each erasure burst is $z$ or fewer time steps in length, and consecutive bursts are separated by a gap of \m{$d-z$} or more unerased time steps, i.e.,
\begin{align*}
\EEE_n^\mB
\triangleq \Big\{
& E\subseteq T_n:
\\*
& (t{\in}E \;\wedge\; t{+}1{\notin}E)
\Rightarrow
|E\cap\{t{+}1,\ldots,t{+}d{-}z\}|=0,
\\*
& (t{\notin}E \;\wedge\; t{+}1{\in}E)
\Rightarrow
|E\cap\{t{+}1,\ldots,t{+}z{+}1\}|\leq z
\Big\}.
\end{align*}
The objective is to construct a code that allows all $n$ messages \m{$\{1,\ldots,n\}$} to be decoded by their respective deadlines under any erasure pattern \m{$E\in\EEE_n^\mB$}.
Let $s_n^\mB$ be the maximum message size that can be achieved by such a code, for a given choice of \m{$(n,c,d,z)$}.

Using the proof technique of Theorem~\ref{thm:theorem:SlidingWindowOptimalCode}, we can show that the constructed intrasession code is asymptotically optimal when $d$ is a multiple of $c$, or when the maximum erasure burst length~$z$ is sufficiently short or long:

\begin{thm:theorem}
\label{thm:theorem:BurstyOptimalCode}
If
\begin{enumerate}
\item
$d$ is a multiple of $c$, or
\item
$d$ is not a multiple of $c$ and
\m{$z\leq c-r_{d,c}$}, or
\item
$d$ is not a multiple of $c$ and
\m{$z\geq d-r_{d,c}=q_{d,c}\,c$},
\end{enumerate}
then the code constructed in Section~\ref{sec:CodeConstruction} is asymptotically optimal in the following sense:
the code achieves a message size of
\[
\sum_{j=1}^{d-z} y_j,
\]
which is equal to the asymptotic maximum achievable message size
$\lim_{n\rightarrow\infty} s_n^\mB$.
\end{thm:theorem}

When the maximum erasure burst length~$z$ takes on intermediate values, intersession coding may become necessary.
We are currently studying optimal convolutional codes for this case.

\appendices

\section{Code Properties}
\label{sec:CodeProperties}

The first property describes when it is possible to decode each message:

\begin{thm:lemma}[Achievability]
\label{thm:lemma:Achievability}
Consider the code constructed in Section~\ref{sec:CodeConstruction} for a given choice of \m{$(c,d)$}.
If message size $s$ satisfies the inequality
\[
s\leq\sum_{j=1}^{\ell} y_j,
\]
where \m{$\yy=(y_1,\ldots,y_d)$} is as defined in Theorem~\ref{thm:theorem:CodingWindowOptimalCode},
then each message~\m{$k\in\ZZ^+$} can be decoded from the data at any $\ell$ time steps in its coding window $W_k$.
\end{thm:lemma}

\noindent
Note that the maximum message size~$s$ that can be supported by the code is
given by
\m{$\sum_{j=1}^{d} y_j=c$}, which corresponds to the choice of \m{$\ell=d$}.

The second property describes a way of partitioning time steps into sets with certain specific properties, which are used in our specification of the worst-case erasure patterns:

\begin{thm:lemma}[Partition of Coding Windows]
\label{thm:lemma:PartitionCodingWindows}
Consider the code constructed in Section~\ref{sec:CodeConstruction} for a given choice of \m{$(c,d)$}.
Consider the first $n$ messages \m{$\{1,\ldots,n\}$}, and the union of their (overlapping) coding windows $T_n$.
The set of time steps~$T_n$ can be partitioned into $d$ sets $T_n^{(1)},\ldots,T_n^{(d)}$, given by
\begin{align*}
T_n^{(i)}
\triangleq
\begin{cases}
\Big\{
\big(j(q_{d,c}+1)+q_{i,c}\big)c+r_{i,c} \in T_n:
j\in\ZZ^+_0
\Big\}
\\
\hspace{15.5em} \text{if } r_{i,c}\leq r_{d,c},
\\[0.75em]
\Big\{
\big(j\,q_{d,c}+q_{i,c}\big)c+r_{i,c} \hspace{2.3em}\in T_n:
j\in\ZZ^+_0
\Big\}
\\
\hspace{15.5em} \text{if } r_{i,c}>r_{d,c},
\end{cases}
\end{align*}
with the following properties:
\begin{enumerate}
\item
\label{item:PartitionProperty1}
Over the time steps in the set $T_n^{(i)}$, each message \m{$k\in\{1,\ldots,n\}$} is allocated $\frac{1}{q_{d,c}+1}$ amount of link bandwidth if \m{$r_{i,c}\leq r_{d,c}$}, and $\frac{1}{q_{d,c}}$ amount of link bandwidth if \m{$r_{i,c}>r_{d,c}$}.
\item
\label{item:PartitionProperty2}
The allocated link bandwidth in $T_n^{(i)}$ for each message \m{$k\in\{1,\ldots,n\}$} is contained within a single time step in its coding window $W_k$
(as opposed to being spread over multiple time steps or being outside of the coding window).
\item
\label{item:PartitionProperty3}
The total amount of link bandwidth over all time steps in $T_n^{(i)}$, i.e., $\big|T_n^{(i)}\big|$, has the following upper bound:
\begin{align*}
\big|T_n^{(i)}\big|
< \begin{cases}
\displaystyle
\frac{n}{q_{d,c}+1}+2
& \text{if } r_{i,c}\leq r_{d,c},
\\[1.5em]
\displaystyle
\frac{n}{q_{d,c}}+2
& \text{if } r_{i,c}>r_{d,c}.
\end{cases}
\end{align*}
\end{enumerate}
\end{thm:lemma}
\begin{figure}
\centering
\subfloat[$(c,d)=(3,9)$]
{\includegraphics[width=250pt]{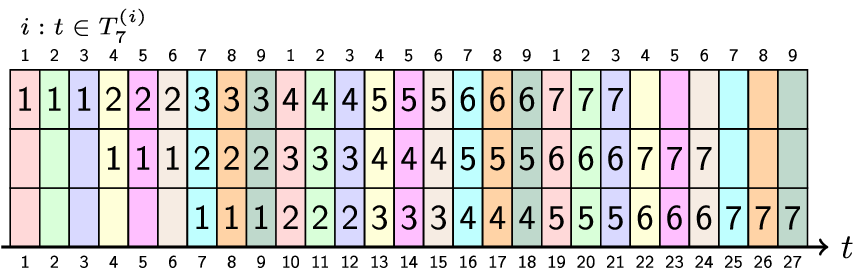}}
\\
\subfloat[$(c,d)=(3,8)$]
{\includegraphics[width=250pt]{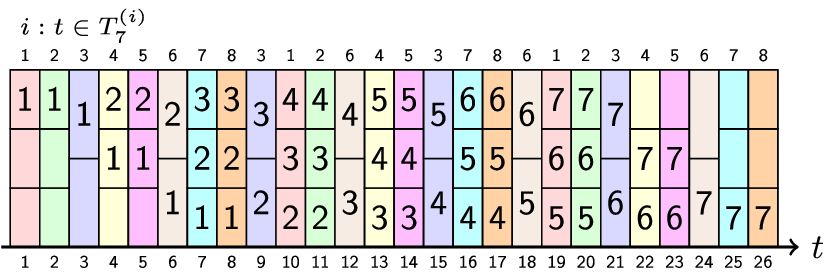}}
\caption{Partitioning of the set of time steps~$T_n$ into the $d$ sets $T_n^{(1)},\ldots,T_n^{(d)}$, in the encoding of messages \m{$\{1,\ldots,7\}$}, for (a)~\m{$(c,d)=(3,9)$} and (b)~\m{$(c,d)=(3,8)$}.
Each set~$T_n^{(i)}$ is assigned a unique color.
The number~$i$ at the top of each time step~$t$ indicates the set~$T_n^{(i)}$ to which $t$ belongs.}
\label{fig:CodePartition}
\end{figure}

\noindent
Fig.~\ref{fig:CodePartition} shows how the set of time steps~$T_n$ is partitioned into the $d$ sets $T_n^{(1)},\ldots,T_n^{(d)}$, for two instances of $(c,d)$.

\section{Proofs of Theorems}
\label{sec:ProofsOfTheorems}

\begin{IEEEproof}[Proof of Lemma~\ref{thm:lemma:Achievability}]
Consider a given message~\m{$k\in\ZZ^+$} and its coding window
\[
W_k
=\big\{(k-1)c+i:i\in\{1,\ldots,d\}\big\}.
\]
At each time step~\m{$t\in W_k$}, message~$k$ is allocated either $\frac{1}{q_{d,c}+1}$ or $\frac{1}{q_{d,c}}$ amount of link bandwidth;
at all other time steps~\m{$t\notin W_k$}, message~$k$ is allocated zero link bandwidth.

Let $x_i$ be the amount of link bandwidth at time step $t=$ \m{$(k-1)c+i$} that is allocated to message~$k$.
Writing $t$ in terms of $q_{i,c}$ and $r_{i,c}$ produces
\[
t
= (k-1)c+i
= \underbrace{(k-1+q_{i,c})}_{q_{t,c}}c+\underbrace{r_{i,c}}_{r_{t,c}}.
\]
It follows from the code construction that the value of $x_i$ depends on $r_{i,c}$;
specifically, two cases are possible:

\m{\textit{Case 1}:}
If \m{$r_{i,c}\leq r_{d,c}$}, then \m{$x_i=\frac{1}{q_{d,c}+1}$}.
Since \m{$i\in\{1,\ldots,d\}$}, this condition corresponds to the case where
\m{$q_{i,c}\in\{0,\ldots,q_{d,c}\}$} and \m{$r_{i,c}\in\{1,\ldots,r_{d,c}\}$}.
Therefore, message~$k$ is allocated $\frac{1}{q_{d,c}+1}$ amount of link bandwidth per time step for a total of \m{$(q_{d,c}+1)r_{d,c}$} time steps in the coding window $W_k$.

\m{\textit{Case 2}:}
If \m{$r_{i,c}>r_{d,c}$}, then \m{$x_i=\frac{1}{q_{d,c}}$}.
Since \m{$i\in\{1,\ldots,d\}$}, this condition corresponds to the case where
\m{$q_{i,c}\in\{0,\ldots,q_{d,c}-1\}$} and \m{$r_{i,c}\in\{r_{d,c}+1,\ldots,c\}$}.
Therefore, message~$k$ is allocated $\frac{1}{q_{d,c}}$ amount of link bandwidth per time step for a total of \m{$q_{d,c}(c-r_{d,c})$} time steps in the coding window $W_k$.

Observe that $\yy$ is simply a vector containing the elements of $\{x_i\}_{i=1}^{d}$ sorted in ascending order.
Since
\[
\sum_{i\in U} x_i
\geq \sum_{j=1}^{|U|} y_j
\quad \forall\; U\subseteq\{1,\ldots,d\},
\]
it follows that over any $\ell$ time steps in the coding window $W_k$, the total amount of link bandwidth allocated to message~$k$ is at least \m{$\sum_{j=1}^{\ell} y_j$}.
Therefore, as long as the message size~$s$ does not exceed \m{$\sum_{j=1}^{\ell} y_j$}, message~$k$ can always be decoded from the data at any $\ell$ time steps in $W_k$.
\end{IEEEproof}

\begin{IEEEproof}[Proof of Lemma~\ref{thm:lemma:PartitionCodingWindows}]
The stated partition can be constructed by assigning each time step \m{$t\in T_n$} to the set~$T_n^{(q_{i,c} c+r_{i,c})}$, where
\begin{align*}
r_{i,c}
&= r_{t,c},
\\
q_{i,c}
&=
\begin{cases}
q_{t,c}-\floor{\frac{q_{t,c}}{q_{d,c}+1}}(q_{d,c}+1)
& \text{if } r_{t,c}\leq r_{d,c},
\\[0.5em]
q_{t,c}-\floor{\frac{q_{t,c}}{q_{d,c}}}q_{d,c}
& \text{if } r_{t,c}>r_{d,c}.
\end{cases}
\end{align*}
Note that index \m{$q_{i,c}\,c+r_{i,c}\in\{1,\ldots,d\}$} since
\m{$q_{i,c}\in\{0,\ldots,q_{d,c}\}$} when \m{$r_{i,c}\in\{1,\ldots,r_{d,c}\}$}, and
\m{$q_{i,c}\in\{0,\ldots,q_{d,c}-1\}$} when \m{$r_{i,c}\in\{r_{d,c}+1,\ldots,c\}$}.
To prove the required code properties, we consider two separate cases:

\m{\textit{Case 1}:}
Consider the set $T_n^{(i)}$ for a choice of $i$ satisfying \m{$r_{i,c}\leq r_{d,c}$}.
Since each time step \m{$t\in T_n^{(i)}$} can be expressed as
\[
t
= \underbrace{\big(j(q_{d,c}+1)+q_{i,c}\big)}_{q_{t,c}}c+\underbrace{r_{i,c}}_{r_{t,c}}
\triangleq t_j,
\quad
\text{where } j\in\ZZ^+_0,
\]
it follows from the code construction that the set of active messages at each time step contains \m{$q_{d,c}+1$} messages, and is given by
\[
A_{t_j}
= \Big\{
\underbrace{j(q_{d,c}{+}1){+}q_{i,c}}_{q_{t,c}}{+}1{-}q_{d,c},
\ldots,
\underbrace{j(q_{d,c}{+}1){+}q_{i,c}}_{q_{t,c}}{+}1
\Big\}.
\]
The smallest time step in $T_n^{(i)}$ corresponds to the choice of \m{$j=0$}, which produces \m{$t_0=q_{i,c}\,c+r_{i,c}=i$} and the set of active messages
\[
A_{t_0}
= \{q_{i,c}+1-q_{d,c},\ldots,q_{i,c}+1\}.
\]
Note that $A_{t_0}$ contains message~$1$ since
\m{$q_{i,c}\in\{0,\ldots,q_{d,c}\}$}, which implies that
\[
q_{i,c}+1-q_{d,c}\leq 1 \leq q_{i,c}+1.
\]
At the other extreme, let the largest time step in $T_n^{(i)}$ correspond to the choice of \m{$j=j'$};
we therefore have
\begin{align}
t_{j'}
\leq (n-1)c+d
< t_{j'+1},
\label{eq:Case1Tj}
\end{align}
and the final set of active messages
\[
A_{t_{j'}}
= \{j'(q_{d,c}+1)+q_{i,c}+1-q_{d,c},\ldots,j'(q_{d,c}+1)+q_{i,c}+1\}.
\]
From the first inequality of \eqref{eq:Case1Tj}, we obtain
\begin{align}
& \big(j'(q_{d,c}+1)+q_{i,c}\big)c+r_{i,c}
\leq (n-1+q_{d,c})c+r_{d,c}
\notag
\\
&\Longrightarrow
n \geq \ceil{\frac{\big(j'(q_{d,c}+1)+q_{i,c}+1-q_{d,c}\big)c+r_{i,c}-r_{d,c}}{c}}
\notag
\\
&\hspace{2.7em}= j'(q_{d,c}+1)+q_{i,c}+1-q_{d,c}+\ceil{\frac{r_{i,c}-r_{d,c}}{c}}
\notag
\\
&\hspace{2.7em}= j'(q_{d,c}+1)+q_{i,c}+1-q_{d,c},
\label{eq:Case1NLower}
\end{align}
where the final step follows from the fact that \m{$1\leq r_{i,c}\leq r_{d,c}\leq c$}, which implies that
\[
-1
< \frac{1-c}{c}
\leq \frac{r_{i,c}-r_{d,c}}{c}
\leq 0
\quad\Longrightarrow\quad \ceil{\frac{r_{i,c}-r_{d,c}}{c}}=0.
\]
From the second inequality of \eqref{eq:Case1Tj}, we obtain
\begin{align}
& (n-1+q_{d,c})c+r_{d,c}
\leq \big((j'{+}1)(q_{d,c}{+}1)+q_{i,c}\big)c+r_{i,c}-1
\notag
\\
&\Longrightarrow
n \leq \floor{\frac{\big((j'{+}1)(q_{d,c}{+}1){+}q_{i,c}{+}1{-}q_{d,c}\big)c
+r_{i,c}{-}r_{d,c}{-}1}{c}}
\notag
\\
&\hspace{2.7em}= (j'{+}1)(q_{d,c}{+}1)+q_{i,c}+1-q_{d,c}
+\floor{\frac{r_{i,c}{-}r_{d,c}{-}1}{c}}
\notag
\\
&\hspace{2.7em}= j'(q_{d,c}+1)+q_{i,c}+1,
\label{eq:Case1NUpper}
\end{align}
where the final step follows from the fact that \m{$1\leq r_{i,c}\leq r_{d,c}\leq c$}, which implies that
\begin{align*}
& -1
= \frac{1-c-1}{c}
\leq \frac{r_{i,c}-r_{d,c}-1}{c}
\leq -\frac{1}{c}
< 0
\\
&\hspace{10em} \Longrightarrow\quad \floor{\frac{r_{i,c}-r_{d,c}-1}{c}}=-1.
\end{align*}
By combining inequalities \eqref{eq:Case1NLower} and \eqref{eq:Case1NUpper}, we arrive at
\[
j'(q_{d,c}+1)+q_{i,c}+1-q_{d,c}
\leq n
\leq j'(q_{d,c}+1)+q_{i,c}+1,
\]
which enables us to infer that $A_{t_{j'}}$ contains message~$n$.

\noindent
For any pair of consecutive time steps \m{$t_j,t_{j+1}\in T_n^{(i)}$}, where
\begin{align*}
t_j &= \big(j(q_{d,c}+1)+q_{i,c}\big)c+r_{i,c},
\\
t_{j+1} &= \big((j+1)(q_{d,c}+1)+q_{i,c}\big)c+r_{i,c},
\end{align*}
we observe that the smallest message in $A_{t_{j+1}}$ is exactly one larger than the largest message in $A_{t_j}$, i.e.,
\begin{align*}
& (j+1)(q_{d,c}+1)+q_{i,c}+1-q_{d,c}
\\
&\hspace{1em}= j(q_{d,c}+1)+q_{i,c}+1-q_{d,c}+q_{d,c}+1
\\
&\hspace{1em}= \big(j(q_{d,c}+1)+q_{i,c}+1\big)+1.
\end{align*}
Thus, there are no overlapping or omitted messages among the sets of active messages corresponding to $T_n^{(i)}$.
Properties~\ref{item:PartitionProperty1} and \ref{item:PartitionProperty2} therefore follow.

\noindent
The total amount of link bandwidth over all time steps in $T_n^{(i)}$, i.e., $\big|T_n^{(i)}\big|$, can be computed by summing over the link bandwidth allocated to the $n$ messages, and adding the unused link bandwidth in the smallest time step (which is allocated to nonpositive dummy messages) and in the largest time step (which is allocated to messages larger than $n$);
this produces the required upper bound of Property~\ref{item:PartitionProperty3}.

\m{\textit{Case 2}:}
Consider the set $T_n^{(i)}$ for a choice of $i$ satisfying \m{$r_{i,c}>r_{d,c}$}.
Since each time step \m{$t\in T_n^{(i)}$} can be expressed as
\[
t
= \underbrace{(j\,q_{d,c}+q_{i,c})}_{q_{t,c}}c+\underbrace{r_{i,c}}_{r_{t,c}}
\triangleq t_j,
\quad
\text{where } j\in\ZZ^+_0,
\]
it follows from the code construction that the set of active messages at each time step contains $q_{d,c}$ messages, and is given by
\[
A_{t_j}
= \Big\{\underbrace{j\,q_{d,c}+q_{i,c}}_{q_{t,c}}+1-(q_{d,c}-1),\ldots,\underbrace{j\,q_{d,c}+q_{i,c}}_{q_{t,c}}+1\Big\}.
\]
The smallest time step in $T_n^{(i)}$ corresponds to the choice of \m{$j=0$}, which produces \m{$t_0=q_{i,c}\,c+r_{i,c}=i$} and the set of active messages
\[
A_{t_0}
= \{q_{i,c}+1-(q_{d,c}-1),\ldots,q_{i,c}+1\}.
\]
Note that $A_{t_0}$ contains message~$1$ since \m{$q_{i,c}\in\{0,\ldots,q_{d,c}-1\}$}, and therefore
\[
q_{i,c}+1-(q_{d,c}-1)\leq 1 \leq q_{i,c}+1.
\]
At the other extreme, let the largest time step in $T_n^{(i)}$ correspond to the choice of \m{$j=j'$};
we therefore have
\begin{align}
t_{j'}
\leq (n-1)c+d
< t_{j'+1},
\label{eq:Case2Tj}
\end{align}
and the final set of active messages
\[
A_{t_{j'}}
= \{j'\,q_{d,c}+q_{i,c}+1-(q_{d,c}-1),\ldots,j'\,q_{d,c}+q_{i,c}+1\}.
\]
From the first inequality of \eqref{eq:Case2Tj}, we obtain
\begin{align}
& (j'\,q_{d,c}+q_{i,c})c+r_{i,c}
\leq (n-1+q_{d,c})c+r_{d,c}
\notag
\\
&\Longrightarrow n \geq \ceil{\frac{(j'\,q_{d,c}+q_{i,c}+1-q_{d,c})c+r_{i,c}-r_{d,c}}{c}}
\notag
\\
&\hspace{2.7em}= j'\,q_{d,c}+q_{i,c}+1-q_{d,c}+\ceil{\frac{r_{i,c}-r_{d,c}}{c}}
\notag
\\
&\hspace{2.7em}= j'\,q_{d,c}+q_{i,c}+1-(q_{d,c}-1),
\label{eq:Case2NLower}
\end{align}
where the final step follows from the fact that \m{$1\leq r_{d,c}< r_{i,c}\leq c$}, which implies that
\[
0
< \frac{r_{i,c}-r_{d,c}}{c}
\leq \frac{c-1}{c}
<1
\quad\Longrightarrow\quad \ceil{\frac{r_{i,c}-r_{d,c}}{c}}=1.
\]
From the second inequality of \eqref{eq:Case2Tj}, we obtain
\begin{align}
& (n-1+q_{d,c})c+r_{d,c}
\leq \big((j'+1)q_{d,c}+q_{i,c}\big)c+r_{i,c}-1
\notag
\\
&\Longrightarrow n \leq \floor{\frac{\big((j'{+}1)q_{d,c}+q_{i,c}+1-q_{d,c}\big)c+r_{i,c}-r_{d,c}-1}{c}}
\notag
\\
&\hspace{2.7em}= (j'+1)q_{d,c}+q_{i,c}+1-q_{d,c}+\floor{\frac{r_{i,c}-r_{d,c}-1}{c}}
\notag
\\
&\hspace{2.7em}= j'\,q_{d,c}+q_{i,c}+1,
\label{eq:Case2NUpper}
\end{align}
where the final step follows from the fact that \m{$1\leq r_{d,c}< r_{i,c}\leq c$}, which implies that
\begin{align*}
& 0
= \frac{1-1}{c}
\leq \frac{r_{i,c}-r_{d,c}-1}{c}
\leq \frac{c-1-1}{c}
< 1
\\
&\hspace{10em} \Longrightarrow\quad \floor{\frac{r_{i,c}-r_{d,c}-1}{c}}=0.
\end{align*}
By combining inequalities \eqref{eq:Case2NLower} and \eqref{eq:Case2NUpper}, we arrive at
\[
j'\,q_{d,c}+q_{i,c}+1-(q_{d,c}-1)
\leq n
\leq j'\,q_{d,c}+q_{i,c}+1,
\]
which enables us to infer that $A_{t_{j'}}$ contains message~$n$.

\noindent
For any pair of consecutive time steps \m{$t_j,t_{j+1}\in T_n^{(i)}$}, where
\begin{align*}
t_j &= \big(j\,q_{d,c}+q_{i,c}\big)c+r_{i,c},
\\
t_{j+1} &= \big((j+1)\,q_{d,c}+q_{i,c}\big)c+r_{i,c},
\end{align*}
we observe that the smallest message in $A_{t_{j+1}}$ is exactly one larger than the largest message in $A_{t_j}$, i.e.,
\begin{align*}
& (j+1)q_{d,c}+q_{i,c}+1-(q_{d,c}-1)
\\
&\hspace{1em}= j\,q_{d,c}+q_{i,c}+1-(q_{d,c}-1)+q_{d,c}
\\
&\hspace{1em}= \big(j\,q_{d,c}+q_{i,c}+1\big)+1.
\end{align*}
Thus, there are no overlapping or omitted messages among the sets of active messages corresponding to $T_n^{(i)}$.
Properties~\ref{item:PartitionProperty1} and \ref{item:PartitionProperty2} therefore follow.

\noindent
The total amount of link bandwidth over all time steps in $T_n^{(i)}$, i.e., $\big|T_n^{(i)}\big|$, can be computed by summing over the link bandwidth allocated to the $n$ messages, and adding the unused link bandwidth in the smallest time step (which is allocated to nonpositive dummy messages) and in the largest time step (which is allocated to messages larger than $n$);
this produces the required upper bound of Property~\ref{item:PartitionProperty3}.
\end{IEEEproof}

\begin{IEEEproof}[Proof of Theorem~\ref{thm:theorem:CodingWindowOptimalCode}]
Consider the code constructed in Section~\ref{sec:CodeConstruction} for a given choice of \m{$(c,d)$}.
According to Lemma~\ref{thm:lemma:Achievability}, if message size~$s$ satisfies the inequality
\[
s
\leq \sum_{j=1}^{d-z} y_j,
\]
then each message \m{$k\in\{1,\ldots,n\}$} can be decoded from the data at any \m{$d-z$} time steps in its coding window~$W_k$.
Therefore, the code achieves a message size of \m{$\sum_{j=1}^{d-z} y_j$}, by allowing all $n$ messages \m{$\{1,\ldots,n\}$} to be decoded by their respective deadlines as long as there are $z$ or fewer erasures in each coding window $W_k$, or equivalently, under any erasure pattern \m{$E\in\EEE_n^\mCW$}.
To demonstrate the asymptotic optimality of the code, we will show that this message size matches the maximum achievable message size~$s_n^\mCW$ in the limit, i.e.,
\begin{align}
\lim_{n\rightarrow\infty} s_n^\mCW
= \sum_{j=1}^{d-z} y_j.
\label{eq:SNCWLimit}
\end{align}

To obtain an upper bound for $s_n^\mCW$, we consider the cut-set bound corresponding to a specific erasure pattern $E'$ from $\EEE_n^\mCW$.
Let \m{$\{1,\ldots,d\}$} be partitioned into two sets $V^{(1)}$ and $V^{(2)}$, where
\begin{align*}
V^{(1)}
&\triangleq \big\{i\in\{1,\ldots,d\}: r_{i,c}\leq r_{d,c}\big\},
\\
V^{(2)}
&\triangleq \big\{i\in\{1,\ldots,d\}: r_{i,c}>r_{d,c}\big\}.
\end{align*}
Let \m{$\vv=(v_1,\ldots,v_d)$} be defined as
\m{$\vv\triangleq\big(\vv^{(1)}\;|\;\vv^{(2)}\big)$}, where
$\vv^{(1)}$ is the vector containing the \m{$(q_{d,c}+1)r_{d,c}$} elements of $V^{(1)}$ sorted in ascending order, and
$\vv^{(2)}$ is the vector containing the \m{$q_{d,c}(c-r_{d,c})$} elements of $V^{(2)}$ sorted in ascending order.
Define the erasure pattern \m{$E'\subseteq T_n$} as follows:
\[
E'
\triangleq \bigcup_{j=d-z+1}^{d} T_n^{(v_j)},
\]
where $T_n^{(i)}$ is as defined in Lemma~\ref{thm:lemma:PartitionCodingWindows}.
The erased time steps in $E'$ have been chosen to coincide with
the larger blocks allocated to each message in the constructed
code.
To show that $E'$ is an admissible erasure pattern, we introduce the following lemma:

\begin{thm:lemma}
\label{thm:lemmma:TnWkIntersection}
If \m{$A\subseteq\{1,\ldots,d\}$}, then
\begin{align}
\Bigg|
\Bigg(\bigcup_{i\in A} T_n^{(i)}\Bigg) \cap W_k
\Bigg|=|A|
\quad \forall\; k\in\{1,\ldots,n\},
\label{eq:TnWkA}
\end{align}
where $T_n^{(i)}$ is as defined in Lemma~\ref{thm:lemma:PartitionCodingWindows}.
\end{thm:lemma}

\begin{IEEEproof}[Proof of Lemma~\ref{thm:lemmma:TnWkIntersection}]
Since the code constructed in Section~\ref{sec:CodeConstruction} allocates a positive amount of link bandwidth to each message \m{$k\in\{1,\ldots,n\}$} at every time step in its coding window~$W_k$, it follows from Property~\ref{item:PartitionProperty2} of Lemma~\ref{thm:lemma:PartitionCodingWindows} that for each \m{$i\in\{1,\ldots,d\}$}, we have
\[
\big|T_n^{(i)}\cap W_k\big|=1
\quad \forall\; k\in\{1,\ldots,n\}.
\]
Equation~\eqref{eq:TnWkA} therefore follows from the fact that $T_n^{(1)},\ldots,T_n^{(d)}$ are disjoint sets.
\end{IEEEproof}

\noindent
Applying Lemma~\ref{thm:lemmma:TnWkIntersection} with \m{$A=\{v_j\}_{j=d-z+1}^{d}$} produces
\[
\big|E'\cap W_k\big|=z
\quad \forall\; k\in\{1,\ldots,n\},
\]
and thus $E'$ is an admissible erasure pattern, i.e., \m{$E'\in\EEE_n^\mCW$}.

Now, consider a code that achieves the maximum message size $s_n^\mCW$.
Such a code must allow all $n$ messages \m{$\{1,\ldots,n\}$} to be decoded under the specific erasure pattern~$E'$.
We therefore have the following cut-set bound for $s_n^\mCW$:
\[
n\,s_n^\mCW
\leq \big|T_n\backslash E'\big|
\;\Longleftrightarrow\;
s_n^\mCW
\leq \frac{1}{n} \big|T_n\backslash E'\big|
= \frac{1}{n} \sum_{j=1}^{d-z} \big|T_n^{(v_j)}\big|.
\]
Applying the upper bounds in Property~\ref{item:PartitionProperty3} of Lemma~\ref{thm:lemma:PartitionCodingWindows}, and writing the resulting expression in terms of $y_j$ produces
\[
s_n^\mCW
\leq \frac{1}{n} \sum_{j=1}^{d-z} \big|T_n^{(v_j)}\big|
\leq \frac{1}{n} \sum_{j=1}^{d-z} (n\,y_j+2).
\]
Since a message size of \m{$\sum_{j=1}^{d-z} y_j$} is known to be achievable (by the constructed code), we have the following upper and lower bounds for $s_n^\mCW$:
\[
\sum_{j=1}^{d-z} y_j
\leq s_n^\mCW
\leq \frac{1}{n} \sum_{j=1}^{d-z} (n\,y_j+2).
\]
These turn out to be matching bounds in the limit as \m{$n\rightarrow\infty$}:
\[
\sum_{j=1}^{d-z} y_j
\leq \lim_{n\rightarrow\infty} s_n^\mCW
\leq \lim_{n\rightarrow\infty} \frac{1}{n} \sum_{j=1}^{d-z} (n\,y_j+2)
= \sum_{j=1}^{d-z} y_j.
\]
We therefore have \eqref{eq:SNCWLimit} as required.
\end{IEEEproof}

\begin{IEEEproof}[Proof of Theorem~\ref{thm:theorem:SlidingWindowOptimalCode}]
Consider the code constructed in Section~\ref{sec:CodeConstruction} for a given choice of \m{$(c,d)$}.
According to Lemma~\ref{thm:lemma:Achievability}, if message size~$s$ satisfies the inequality
\[
s
\leq \sum_{j=1}^{d-z} y_j,
\]
then each message \m{$k\in\{1,\ldots,n\}$} can be decoded from the data at any \m{$d-z$} time steps in its coding window~$W_k$.
Therefore, the code achieves a message size of \m{$\sum_{j=1}^{d-z} y_j$}, by allowing all $n$ messages \m{$\{1,\ldots,n\}$} to be decoded by their respective deadlines as long as there are $z$ or fewer erasures in each coding window $W_k$, which is indeed the case when there are $z$ or fewer erasures in each sliding window $L_t$, or equivalently, under any erasure pattern \m{$E\in\EEE_n^\mSW$}.
To demonstrate the asymptotic optimality of the code, we will show that this message size matches the maximum achievable message size~$s_n^\mSW$ in the limit, i.e.,
\begin{align}
\lim_{n\rightarrow\infty} s_n^\mSW
= \sum_{j=1}^{d-z} y_j.
\label{eq:SNSWLimit}
\end{align}
We consider two cases separately, depending on whether $d$ is a multiple of $c$:

\m{\textit{Case 1}:}
Suppose that $d$ is a multiple of $c$.
In this case, the message size achieved by the constructed code simplifies to
\[
\sum_{j=1}^{d-z} y_j
= \frac{d-z}{q_{d,c}+1}
= \frac{d-z}{d}c.
\]

To obtain an upper bound for $s_n^\mSW$, we consider the cut-set bound corresponding to a specific periodic erasure pattern \m{$E'\subseteq T_n$} given by
\[
E'
\triangleq \big\{j\,d+i\in T_n:j\in\ZZ_0^+,i\in\{1,\ldots,z\}\big\}.
\]
Since $E'$ comprises alternating intervals of $z$ erased time steps and \m{$d-z$} unerased time steps, we have exactly $z$ erasures in each sliding window $L_t$;
therefore, $E'$ is an admissible erasure pattern, i.e., \m{$E'\in\EEE_n^\mSW$}.

Now, consider a code that achieves the maximum message size $s_n^\mSW$.
Such a code must allow all $n$ messages \m{$\{1,\ldots,n\}$} to be decoded under the specific erasure pattern~$E'$.
We therefore have the following cut-set bound for $s_n^\mSW$:
\[
n\,s_n^\mSW
\leq \big|T_n\backslash E'\big|
= \floor{\frac{(n-1)c+d}{d}}(d-z)+\max(r'-z,0),
\]
where
\[
r'
\triangleq (n-1)c+d-\floor{\frac{(n-1)c+d}{d}}d.
\]
Further simplification produces
\[
s_n^\mSW
\leq \frac{1}{n} \frac{(n-1)c+2d}{d} (d-z)
= \frac{d-z}{d} \left(c+\frac{2d-c}{n}\right).
\]
Since a message size of \m{$\frac{d-z}{d}c$} is known to be achievable (by the constructed code), we have the following upper and lower bounds for $s_n^\mSW$:
\[
\frac{d-z}{d}c
\leq s_n^\mSW
\leq \frac{d-z}{d} \left(c+\frac{2d-c}{n}\right).
\]
These turn out to be matching bounds in the limit as \m{$n\rightarrow\infty$}:
\[
\frac{d-z}{d}c
\leq \lim_{n\rightarrow\infty} s_n^\mSW
\leq \lim_{n\rightarrow\infty} \frac{d-z}{d} \left(c+\frac{2d-c}{n}\right)
= \frac{d-z}{d}c.
\]
We therefore have \eqref{eq:SNSWLimit} as required.

\m{\textit{Case 2}:}
Suppose that $d$ is not a multiple of $c$.
Consider a specific \emph{base} erasure pattern \m{$E'\subseteq T_n$} given by
\[
E'
\triangleq \bigcup_{j=d-z+1}^{d} T_n^{(v_j)},
\]
where $T_n^{(i)}$ is as defined in Lemma~\ref{thm:lemma:PartitionCodingWindows}, and \m{$\vv=(v_1,\ldots,v_d)$} is as defined in the proof of Theorem~\ref{thm:theorem:CodingWindowOptimalCode}.
The erased time steps in $E'$ have been chosen to coincide with
the larger blocks allocated to each message in the constructed
code.
From $E'$, we derive the erasure patterns $E'_1,\ldots,E'_n$ given by
\[
E'_k
\triangleq E'\cap W_k
= \bigcup_{j=d-z+1}^{d}
\Big(
T_n^{(v_j)}\cap W_k
\Big).
\]
Applying Lemma~\ref{thm:lemmma:TnWkIntersection} with \m{$A=\{v_j\}_{j=d-z+1}^{d}$} produces
\[
\big|E'_k\big|
= \big|E'\cap W_k\big|=z
\quad \forall\; k\in\{1,\ldots,n\},
\]
which implies that
\[
|E'_k\cap L_t|\leq z
\quad \forall\; t\in\{1,\ldots,(n-1)c+1\}
\]
for each \m{$k\in\{1,\ldots,n\}$}.
Thus, $E'_k$ is an admissible erasure pattern, i.e., \m{$E'_k\in\EEE_n^\mSW$}, for each \m{$k\in\{1,\ldots,n\}$}.

To obtain an upper bound for $s_n^\mSW$, we introduce the following lemma:

\begin{thm:lemma}
\label{thm:lemma:EntropyBound}
Suppose that a code achieves a message size of $s$ under a given set of erasure patterns~$\EEE$ for a given choice of \m{$(n,c,d)$}.
Let $X_t$ be a random variable representing the coded data transmitted at time step \m{$t\in T_n$},
let $M_k$ be a random variable representing message \m{$k\in\{1,\ldots,n\}$}, and
define \m{$X[A]\triangleq(X_t)_{t\in A}$}.
If \m{$E\subseteq T_n$} is such that \m{$E\cap W_k$} is an admissible erasure pattern, i.e., \m{$(E\cap W_k)\in\EEE$}, for each \m{$k\in\{1,\ldots,n\}$}, then for each \m{$k\in\{1,\ldots,n\}$},
\begin{align}
H\Big(
X[W_k\backslash E]
\,\Big|\,
M_1^k,
X_1^{(k-1)c}
\Big)
\leq
\big|T_k\backslash E\big|-k\,s.
\label{eq:EntropyBound}
\end{align}
\end{thm:lemma}

\begin{IEEEproof}[Proof of Lemma~\ref{thm:lemma:EntropyBound}]
We will prove by induction that inequality~\eqref{eq:EntropyBound} holds for any \m{$k\in\{1,\ldots,n\}$}.

(Base case)
Consider the case of \m{$k=1$}.
From the definition of mutual information, we have
\begin{align*}
I\big(X[W_1\backslash E]\,;\,M_1\big)
&=
H\big(X[W_1\backslash E]\big)
- H\big(X[W_1\backslash E]\,\big|\,M_1\big)
\\
&=
H\big(M_1\big)
- H\big(M_1\,\big|\,X[W_1\backslash E]\big).
\end{align*}
Rearranging terms produces
\begin{align}
H\big(X[W_1\backslash E]\,\big|\,M_1\big)
&=
H\big(X[W_1\backslash E]\big)
- H\big(M_1\big)
\notag
\\*
&\hspace{1.2em}
+ H\big(M_1\,\big|\,X[W_1\backslash E]\big).
\label{eq:HXW1EM1}
\end{align}
Since \m{$T_1=W_1$} and \m{$H(X_t)\leq 1$} for any $t$ because of the unit link bandwidth, we have
\begin{align}
H\big(X[W_1\backslash E]\big)
= H\big(X[T_1\backslash E]\big)
\leq \big|T_1\backslash E\big|.
\label{eq:HXW1E}
\end{align}
Furthermore, since \m{$E\cap W_1$} is an admissible erasure pattern, message~$1$ must be decodable from the coded data at time steps
\m{$T_1\backslash(E\cap W_1)$} $=$
\m{$W_1\backslash(E\cap W_1)$} $=$
\m{$W_1\backslash E$}, and so
\begin{align}
H\big(M_1\,\big|\,X[W_1\backslash E]\big) = 0.
\label{eq:HM1}
\end{align}
Substituting \eqref{eq:HXW1E}, \eqref{eq:HM1}, and \m{$H(M_1)=s$} into \eqref{eq:HXW1EM1} yields
\[
H\big(X[W_1\backslash E]\,\big|\,M_1\big)
\leq \big|T_1\backslash E\big|-s,
\]
as required.

(Inductive step)
Suppose that
\begin{align}
H\Big(
X[W_k\backslash E]
\,\Big|\,
M_1^k,
X_1^{(k-1)c}
\Big)
\leq
\big|T_k\backslash E\big|-k\,s
\label{eq:HXWkEInductiveHypothesis}
\end{align}
for some \m{$k\in\{1,\ldots,n-1\}$}.
From the definition of conditional mutual information, we have
\begin{align*}
& I\Big(X[W_{k+1}\backslash E]\,;\,
M_{k+1}\,\Big|\,M_1^k,X_1^{kc}\Big)
\\
&=
H\Big(X[W_{k+1}\backslash E]\,\Big|\,
M_1^k,X_1^{kc}\Big)
\\*
&\hspace{3em}
- H\Big(X[W_{k+1}\backslash E]\,\Big|\,
M_1^{k+1},X_1^{kc}\Big)
\\
&=
H\Big(M_{k+1}\,\Big|\,
M_1^k,X_1^{kc}\Big)
\\*
&\hspace{3em}
- H\Big(M_{k+1}\,\Big|\,
M_1^k,X[\{1,\ldots,kc\} \cup (W_{k+1}\backslash E)]\Big).
\end{align*}
Rearranging terms produces
\begin{align}
& H\Big(X[W_{k+1}\backslash E]\,\Big|\,
M_1^{k+1},X_1^{kc}\Big)
\notag
\\
&=
H\Big(X[W_{k+1}\backslash E]\,\Big|\,
M_1^k,X_1^{kc}\Big)
\notag
\\*
&\hspace{2em}
- H\Big(M_{k+1}\,\Big|\,
M_1^k,X_1^{kc}\Big)
\notag
\\*
&\hspace{2em}
+ H\Big(M_{k+1}\,\Big|\,
M_1^k,X[\{1,\ldots,kc\}\cup(W_{k+1}\backslash E)]\Big).
\label{eq:HXWk1Ek1}
\end{align}
Since messages are independent and message \m{$k+1$} is created at time step \m{$kc+1$}, we have
\begin{align}
H\Big(M_{k+1}\,\Big|\,M_1^k,X_1^{kc}\Big)
= H\big(M_{k+1}\big)
= s.
\label{eq:HMk1}
\end{align}
Furthermore, since \m{$E\cap W_{k+1}$} is an admissible erasure pattern, message \m{$k+1$} must be decodable from the coded data at time steps
\m{$T_{k+1}\backslash(E\cap W_{k+1})$} $=$
\m{$(T_{k+1}\backslash W_{k+1})\cup(W_{k+1}\backslash E)$} $=$
\m{$\{1,\ldots,kc\}\cup(W_{k+1}\backslash E)$},
and so
\begin{align}
H\Big(M_{k+1}\,\Big|\,
M_1^k,X[\{1,\ldots,kc\}\cup(W_{k+1}\backslash E)]\Big)
= 0.
\label{eq:HMk1M1kX}
\end{align}
Substituting \eqref{eq:HMk1} and \eqref{eq:HMk1M1kX} into \eqref{eq:HXWk1Ek1} yields
\begin{align}
& H\Big(X[W_{k+1}\backslash E]\,\Big|\,
M_1^{k+1},X_1^{kc}\Big)
\notag
\\
&= H\Big(X[W_{k+1}\backslash E]\,\Big|\,
M_1^k,X_1^{kc}\Big)
- s
\notag
\\
&\overset{\text{(a)}}{\leq}
H\Big(X\big[(W_k\backslash E)\cup
(W_{k+1}\backslash E)\big]\,\Big|\,
M_1^k,X_1^{kc}\Big)
- s
\notag
\\
&\overset{\text{(b)}}{\leq}
H\Big(X\big[(W_k\backslash E)\cup
(W_{k+1}\backslash E)\big]\,\Big|\,
M_1^k,X_1^{(k-1)c}\Big)
- s
\notag
\\
&\overset{\text{(c)}}{\leq}
H\Big(X[W_k\backslash E]\,\Big|\,
M_1^k,X_1^{(k-1)c}\Big)
\notag
\\*
&\hspace{1.5em}
+ H\Big(X\big[(W_{k+1}\backslash E)
\big\backslash(W_k\backslash E)\big]\,\Big)
- s
\notag
\\
&\overset{\text{(d)}}{\leq}
\big|T_k\backslash E\big|-k\,s
+ \Big|(W_{k+1}\backslash E)\big\backslash
(W_k\backslash E)\Big|
- s
\notag
\\
&\overset{\text{(e)}}{=}
\big|T_{k+1}\backslash E\big|-(k+1)s,
\notag
\end{align}
as required, where
\begin{itemize}
\item[(a)]
follows from the addition of random variables
\m{$X[W_k\backslash E]$} in the entropy term;
\item[(b)]
follows from the removal of conditioned random variables \m{$X_{(k-1)c+1}^{kc}$} in the entropy term;
\item[(c)]
follows from the chain rule for joint entropy, and the removal of conditioned random variables
\m{$X[W_k\backslash E]$},
\m{$M_1^k$}, and \m{$X_1^{(k-1)c}$} in the second entropy term;
\item[(d)]
follows from the inductive hypothesis~\eqref{eq:HXWkEInductiveHypothesis}, and the fact that \m{$H(X_t)\leq 1$} for any $t$ because of the unit link bandwidth;
\item[(e)]
follows from the fact that
\begin{align*}
& \big|T_k\backslash E\big|
+\big|(W_{k+1}\backslash E)\big\backslash
(W_k\backslash E)\big|
\\
&\quad = \big|T_k\backslash E\big|
+\big|(W_{k+1}\backslash W_k)\big\backslash E\big|
\\
&\quad = \big|T_k\backslash E\big|
+\big|(T_{k+1}\backslash T_k)\big\backslash E\big|
\\
&\quad = \big|T_{k+1}\backslash E\big|.
\end{align*}
\end{itemize}
\end{IEEEproof}

Applying Lemma~\ref{thm:lemma:EntropyBound} with
\m{$\EEE=\EEE_n^\mSW$} and
\m{$E=E'$} to an optimal code that achieves a message size of $s_n^\mSW$ produces
\[
H\Big(
X[W_k\backslash E']
\,\Big|\,
M_1^k,
X_1^{(k-1)c}
\Big)
\leq
\big|T_k\backslash E'\big|-k\,s_n^\mSW
\]
for any \m{$k\in\{1,\ldots,n\}$}.
Since the conditional entropy term is nonnegative, it follows that for the choice of \m{$k=n$}, we have
\[
\big|T_n\backslash E'\big|-n\,s_n^\mSW
\geq 0
\;\Longleftrightarrow\;
s_n^\mSW
\leq \frac{1}{n} \big|T_n\backslash E'\big|
= \frac{1}{n} \sum_{j=1}^{d-z} \big|T_n^{(v_j)}\big|.
\]
Applying the upper bounds in Property~\ref{item:PartitionProperty3} of Lemma~\ref{thm:lemma:PartitionCodingWindows}, and writing the resulting expression in terms of $y_j$ produces
\[
s_n^\mSW
\leq \frac{1}{n} \sum_{j=1}^{d-z} \big|T_n^{(v_j)}\big|
\leq \frac{1}{n} \sum_{j=1}^{d-z} (n\,y_j+2).
\]
Since a message size of \m{$\sum_{j=1}^{d-z} y_j$} is known to be achievable (by the constructed code), we have the following upper and lower bounds for $s_n^\mSW$:
\[
\sum_{j=1}^{d-z} y_j
\leq s_n^\mSW
\leq \frac{1}{n} \sum_{j=1}^{d-z} (n\,y_j+2).
\]
These turn out to be matching bounds in the limit as \m{$n\rightarrow\infty$}:
\[
\sum_{j=1}^{d-z} y_j
\leq \lim_{n\rightarrow\infty} s_n^\mSW
\leq \lim_{n\rightarrow\infty} \frac{1}{n} \sum_{j=1}^{d-z} (n\,y_j+2)
= \sum_{j=1}^{d-z} y_j.
\]
We therefore have \eqref{eq:SNSWLimit} as required.
\end{IEEEproof}

\begin{IEEEproof}[Proof of Theorem~\ref{thm:theorem:BurstyOptimalCode}]
Observe that under each erasure pattern \m{$E\in\EEE_n^\mB$}, the coding window~$W_k$ for each message \m{$k\in\{1,\ldots,n\}$} contains at most $z$ erasures:
if $W_k$ intersects with zero erasure bursts, then it contains zero erasures;
if $W_k$ intersects with exactly one erasure burst, then it contains at most $z$ erasures, i.e., the maximum length of a burst;
if $W_k$ intersects with two or more erasure bursts, then it contains a gap of at least \m{$d-z$} unerased time steps between consecutive bursts, and therefore contains at most $z$ erasures.

Consider the code constructed in Section~\ref{sec:CodeConstruction} for a given choice of \m{$(c,d)$}.
According to Lemma~\ref{thm:lemma:Achievability}, if message size~$s$ satisfies the inequality
\[
s
\leq \sum_{j=1}^{d-z} y_j,
\]
then each message \m{$k\in\{1,\ldots,n\}$} can be decoded from the data at any \m{$d-z$} time steps in its coding window~$W_k$.
Therefore, the code achieves a message size of \m{$\sum_{j=1}^{d-z} y_j$}, by allowing all $n$ messages \m{$\{1,\ldots,n\}$} to be decoded by their respective deadlines as long as there are $z$ or fewer erasures in each coding window $W_k$, which is indeed the case under any erasure pattern \m{$E\in\EEE_n^\mB$}.
To demonstrate the asymptotic optimality of the code, we will show that this message size matches the maximum achievable message size~$s_n^\mB$ in the limit, i.e.,
\begin{align}
\lim_{n\rightarrow\infty} s_n^\mB
= \sum_{j=1}^{d-z} y_j,
\label{eq:SNBLimit}
\end{align}
for the following three cases:

\m{\textit{Case 1}:}
Suppose that $d$ is a multiple of $c$.
In this case, the message size achieved by the constructed code simplifies to
\[
\sum_{j=1}^{d-z} y_j
= \frac{d-z}{q_{d,c}+1}
= \frac{d-z}{d}c.
\]
To obtain an upper bound for $s_n^\mB$, we consider the cut-set bound corresponding to a specific periodic erasure pattern \m{$E'\subseteq T_n$} given by
\[
E'
\triangleq \big\{j\,d+i\in T_n:j\in\ZZ_0^+,i\in\{1,\ldots,z\}\big\}.
\]
Since $E'$ comprises alternating intervals of $z$ erased time steps and \m{$d-z$} unerased time steps, it is an admissible erasure pattern, i.e., \m{$E'\in\EEE_n^\mB$}.

The rest of the proof leading to the obtainment of \eqref{eq:SNBLimit} is the same as that of Case~1 in the proof of Theorem~\ref{thm:theorem:SlidingWindowOptimalCode}, with $s_n^\mSW$ replaced by $s_n^\mB$.

\m{\textit{Case 2}:}
Suppose that $d$ is not a multiple of $c$, and \m{$z\leq c-r_{d,c}$}.
In this case, the message size achieved by the constructed code simplifies to
\[
\sum_{j=1}^{d-z} y_j
= c-\sum_{j=d-z+1}^{d} y_j
= c-\frac{z}{q_{d,c}}.
\]
Consider a specific \emph{base} erasure pattern \m{$E'\subseteq T_n$} given by
\[
E'
\triangleq \bigcup_{j=d-z+1}^{d} T_n^{(v_j)},
\]
where $T_n^{(i)}$ is as defined in Lemma~\ref{thm:lemma:PartitionCodingWindows}, and \m{$\vv=(v_1,\ldots,v_d)$} is as defined in the proof of Theorem~\ref{thm:theorem:CodingWindowOptimalCode}.
The erased time steps in $E'$ have been chosen to coincide with
the larger blocks allocated to each message in the constructed
code.
In this case, $E'$ simplifies to
\begin{align*}
E'
&= \bigcup_{r_{i,c}=c-z+1}^{c} T_n^{((q_{d,c}-1)c+r_{i,c})}
\\
&= \Big\{
\big((j+1)q_{d,c}-1\big)c+r_{i,c}\in T_n:
\\*[-0.6em]
&\hspace{7.3em}
j\in\ZZ^+_0,
r_{i,c}\in\{c-z+1,\ldots,c\}
\Big\},
\end{align*}
which follows from the definition of $T_n^{(i)}$ and the fact that \m{$r_{i,c}>r_{d,c}$} when
\m{$r_{i,c}\in\{c-z+1,\ldots,c\}$}.
Observe that $E'$ comprises alternating intervals of $z$ erased time steps and \m{$q_{d,c}\,c-z$} unerased time steps, with each interval of erased time steps corresponding to a specific choice of \m{$j\in\ZZ^+_0$}.
Since each erased time step \m{$t\in E'$} can be expressed as
\[
t
= \underbrace{\big((j+1)q_{d,c}-1\big)}_{q_{t,c}}c+\underbrace{r_{i,c}}_{r_{t,c}},
\]
it follows from Section~\ref{sec:CodeConstruction} that the set of active messages $A_t$ at time step~$t$ is given by
\[
A_t
= \Big\{
\underbrace{(j+1)q_{d,c}}_{q_{t,c}+1}-(q_{d,c}-1),\ldots,
\underbrace{(j+1)q_{d,c}}_{q_{t,c}+1}
\Big\}.
\]
Therefore, the set of active messages $A_t$ is the same at every time step~$t$ in a given interval of $z$ erased time steps (corresponding to a specific $j$).

From $E'$, we derive the erasure patterns $E'_1,\ldots,E'_n$ given by
\[
E'_k
\triangleq E'\cap W_k
= \bigcup_{j=d-z+1}^{d}
\Big(
T_n^{(v_j)}\cap W_k
\Big).
\]
Applying Lemma~\ref{thm:lemmma:TnWkIntersection} with \m{$A=\{v_j\}_{j=d-z+1}^{d}$} produces
\[
\big|E'_k\big|
= \big|E'\cap W_k\big|=z
\quad \forall\; k\in\{1,\ldots,n\}.
\]
Let \m{$t'\in E'_k$} be one of the $z$ erased time steps in $W_k$ under erasure pattern~$E'_k$.
As previously established, $t'$ belongs to an interval of $z$ erased time steps in $E'$ that have the same set of active messages $A_{t'}$ (which contains message~$k$).
It follows that this interval of $z$ erased time steps is also in $E'_k$, and must therefore constitute $E'_k$ itself.
Thus, $E'_k$ is an admissible erasure pattern, i.e., \m{$E'_k\in\EEE_n^\mB$}, for each \m{$k\in\{1,\ldots,n\}$}, because it comprises a single erasure burst of $z$ time steps.

Applying Lemma~\ref{thm:lemma:EntropyBound} with
\m{$\EEE=\EEE_n^\mB$} and
\m{$E=E'$} to an optimal code that achieves a message size of $s_n^\mB$ produces
\[
H\Big(
X[W_k\backslash E']
\,\Big|\,
M_1^k,
X_1^{(k-1)c}
\Big)
\leq
\big|T_k\backslash E'\big|-k\,s_n^\mB
\]
for any \m{$k\in\{1,\ldots,n\}$}.
Since the conditional entropy term is nonnegative, it follows that for the choice of \m{$k=n$}, we have
\[
\big|T_n\backslash E'\big|-n\,s_n^\mB
\geq 0
\;\Longleftrightarrow\;
s_n^\mB
\leq \frac{1}{n} \big|T_n\backslash E'\big|
= \frac{1}{n} \sum_{j=1}^{d-z} \big|T_n^{(v_j)}\big|.
\]
The rest of the proof leading to the obtainment of \eqref{eq:SNBLimit} is the same as that of Case~2 in the proof of Theorem~\ref{thm:theorem:SlidingWindowOptimalCode}, with $s_n^\mSW$ replaced by $s_n^\mB$.

\m{\textit{Case 3}:}
Suppose that $d$ is not a multiple of $c$, and \m{$z\geq d-r_{d,c}=q_{d,c}\,c$}.
In this case, the message size achieved by the constructed code simplifies to
\[
\sum_{j=1}^{d-z} y_j
= \frac{d-z}{q_{d,c}+1}.
\]
Consider a specific \emph{base} erasure pattern \m{$E'\subseteq T_n$} given by
\[
E'
\triangleq \bigcup_{j=d-z+1}^{d} T_n^{(v_j)},
\]
where $T_n^{(i)}$ is as defined in Lemma~\ref{thm:lemma:PartitionCodingWindows}, and \m{$\vv=(v_1,\ldots,v_d)$} is as defined in the proof of Theorem~\ref{thm:theorem:CodingWindowOptimalCode}.
The erased time steps in $E'$ have been chosen to coincide with
the larger blocks allocated to each message in the constructed
code.
In this case, $E'$ simplifies to
\begin{align*}
E'
&= T_n\bigg\backslash
\bigg(
\bigcup_{r_{i,c}=1}^{d-z} T_n^{(r_{i,c})}
\bigg)
\\
&= T_n\bigg\backslash
\Big\{
\big(j(q_{d,c}+1)\big)c+r_{i,c}\in T_n:
\\*[-0.6em]
&\hspace{7.8em}
j\in\ZZ^+_0,
r_{i,c}\in\{1,\ldots,d-z\}
\Big\},
\end{align*}
which follows from the definition of $T_n^{(i)}$ and the fact that \m{$r_{i,c}\leq r_{d,c}$} when
\m{$r_{i,c}\in\{1,\ldots,d-z\}$}.
Observe that $E'$ comprises alternating intervals of \m{$d-z$} unerased time steps and \m{$(q_{d,c}+1)\,c-(d-z)$} $=$ \m{$c-r_{d,c}+z$} erased time steps, with each interval of unerased time steps corresponding to a specific choice of \m{$j\in\ZZ^+_0$}.
Since each unerased time step \m{$t\in T_n\backslash E'$} can be expressed as
\[
t
= \underbrace{\big(j(q_{d,c}+1)\big)}_{q_{t,c}}c+\underbrace{r_{i,c}}_{r_{t,c}},
\]
it follows from Section~\ref{sec:CodeConstruction} that the set of active messages $A_t$ at time step~$t$ is given by
\[
A_t
= \Big\{
\underbrace{j(q_{d,c}+1)}_{q_{t,c}}+1-q_{d,c},\ldots,
\underbrace{j(q_{d,c}+1)}_{q_{t,c}}+1
\Big\}.
\]
Therefore, the set of active messages $A_t$ is the same at every time step~$t$ in a given interval of \m{$d-z$} unerased time steps (corresponding to a specific $j$).

From $E'$, we derive the erasure patterns $E'_1,\ldots,E'_n$ given by
\[
E'_k
\triangleq E'\cap W_k
= \bigcup_{j=d-z+1}^{d}
\Big(
T_n^{(v_j)}\cap W_k
\Big).
\]
Applying Lemma~\ref{thm:lemmma:TnWkIntersection} with \m{$A=\{v_j\}_{j=d-z+1}^{d}$} produces
\[
\big|E'_k\big|
= \big|E'\cap W_k\big|=z
\quad \forall\; k\in\{1,\ldots,n\}.
\]
Let \m{$t'\in W_k\backslash E'_k$} be one of the \m{$d-z$} unerased time steps in $W_k$ under erasure pattern~$E'_k$.
As previously established, $t'$ belongs to an interval of \m{$d-z$} unerased time steps in \m{$T_n\backslash E'$} that have the same set of active messages $A_{t'}$ (which contains message~$k$).
It follows that this interval of \m{$d-z$} unerased time steps is also in \m{$W_k\backslash E'_k$}, and must therefore constitute \m{$W_k\backslash E'_k$} itself.
Thus, $E'_k$ is an admissible erasure pattern, i.e., \m{$E'_k\in\EEE_n^\mB$}, for each \m{$k\in\{1,\ldots,n\}$}, because it comprises either a single erasure burst of $z$ time steps, or two erasure bursts with a combined length of $z$ time steps separated by a gap of \m{$d-z$} unerased time steps.

Applying Lemma~\ref{thm:lemma:EntropyBound} with
\m{$\EEE=\EEE_n^\mB$} and
\m{$E=E'$} to an optimal code that achieves a message size of $s_n^\mB$ produces
\[
H\Big(
X[W_k\backslash E']
\,\Big|\,
M_1^k,
X_1^{(k-1)c}
\Big)
\leq
\big|T_k\backslash E'\big|-k\,s_n^\mB
\]
for any \m{$k\in\{1,\ldots,n\}$}.
Since the conditional entropy term is nonnegative, it follows that for the choice of \m{$k=n$}, we have
\[
\big|T_n\backslash E'\big|-n\,s_n^\mB
\geq 0
\;\Longleftrightarrow\;
s_n^\mB
\leq \frac{1}{n} \big|T_n\backslash E'\big|
= \frac{1}{n} \sum_{j=1}^{d-z} \big|T_n^{(v_j)}\big|.
\]
The rest of the proof leading to the obtainment of \eqref{eq:SNBLimit} is the same as that of Case~2 in the proof of Theorem~\ref{thm:theorem:SlidingWindowOptimalCode}, with $s_n^\mSW$ replaced by $s_n^\mB$.
\end{IEEEproof}

\end{document}